\documentclass[pra,superscriptaddress,twocolumn,showpacs]{revtex4}
\usepackage{epsfig}
\usepackage{bm}
\usepackage{amsmath}

\begin{document}

\title{Strong-field approximation for intense-laser
atom processes: the choice of gauge}
\author{D. Bauer}
\affiliation{Max-Planck-Institut f\"ur Kernphysik, Heidelberg, Germany}
\author{D. B. Milo\v{s}evi\'{c}}
\affiliation{Faculty of Science, University of Sarajevo,
Zmaja od Bosne 35, 71000 Sarajevo, Bosnia and Herzegovina}
\author{W. Becker}
\affiliation{Max-Born-Institut, Max-Born-Strasse 2a, 12489 Berlin, Germany}
\date{\today}
\begin{abstract}
The strong-field approximation can be and has been applied in both
length gauge and velocity gauge with quantitatively conflicting
answers. For ionization of negative ions with a ground state of odd
parity, the predictions of the two gauges differ qualitatively: in
the envelope of the angular-resolved  energy spectrum, dips in one
gauge correspond to humps in the other. We show that the
length-gauge SFA matches the exact numerical solution of the
time-dependent Schr\"odinger equation.
\end{abstract}


\newcommand{\vecr}{\mathbf{r}}
\newcommand{\vecA}{\mathbf{A}}
\newcommand{\vecE}{\mathbf{E}}
\newcommand{\veca}{\mathbf{a}}
\newcommand{\vecp}{\mathbf{p}}
\newcommand{\vecv}{\mathbf{v}}
\newcommand{\VS}{\psi^{(\mathrm{Vv})}}
\newcommand{\bq}{\begin{equation}}
\newcommand{\eq}{\end{equation}}
\newcommand{\Hf}{H_{\mathrm{F}x}}
\newcommand{\del}{\mbox{\boldmath{$\nabla$}}}
\newcommand{\vece}{\mathbf{e}}
\newcommand{\Zeff}{Z_{\mbox{\scriptsize eff}}}
\maketitle

Quantum mechanics is gauge invariant: it is easily proven that a
given physical quantity can be evaluated in any gauge with the same
result \cite{ct}. In nonrelativistic quantum mechanics, when
the dipole approximation is adopted, the interaction of an
atom with a time-dependent field such as a laser field is usually
described in either one of two gauges: the length gauge (L gauge) or
the velocity gauge (V gauge). In numerical solutions of the
time-dependent Schr\"odinger equation (TDSE), gauge invariance has been
confirmed many times. In analytical work, however, some
approximations almost always have to be adopted. There is no formal
reason of why after such approximations the resulting theory 
should still be gauge invariant. Indeed, the lack of gauge invariance
after what seems like very well justified approximations has driven
many a researcher to despair \cite{jaynes}.

In this paper, we will address one of the most  glaring
manifestations of this ``gauge problem'': the lack of gauge
invariance of the strong-field approximation (SFA) in
intense-laser--atom physics \cite{KFR}. The SFA underlies almost any analytical
approach to total ionization rates, above-threshold ionization,
high-order harmonic generation, and nonsequential double ionization,
both of atoms and of molecules. Briefly, it assumes that the initial
bound state of the atom or molecule is unaffected by the laser field
while the final state, which is in the continuum, does not feel the
presence of the binding potential. The lack of gauge invariance of
the SFA has been noted many times; see, e.g., Ref.~\cite{SBBS}. 
Comparisons that
have been carried out indeed have exhibited significant
disagreements between the results obtained from L gauge and V gauge
\cite{palermo}. Different authors have preferred different gauges.
The question of which gauge is superior for which problem has often
been raised, but never led to any consensus about its answer. Below,
we will give an answer for the case of a short-range binding potential,
where the SFA is expected to be most accurate \cite{GK}, by comparing the SFA in L gauge and V gauge with the numerical solution of the TDSE.

For a fixed nucleus and in the single-active-electron approximation,
where the effects of all electrons but one are absorbed into an effective
binding potential,
the complete Hamiltonian in the presence of an external electromagnetic
field can be decomposed as
\begin{equation}
H_x(t)= H_0 + H_{\mathrm{I}x}(t), \label{totalH}
\end{equation}
where the subscript $x$ specifies the gauge ($x=\mathrm{L},\mathrm{V}$) and
\begin{equation}
H_0 = \frac{\hat\vecp^2}{2m} + V(\vecr)\ \ \mathrm{with}\ \ \hat\vecp=-i\del.\label{atomH}
\end{equation}
This operator contains the binding potential $V(\vecr)$ and is
independent of the gauge. With the dipole approximation, which
neglects the space dependence of the electric field and the vector
potential, so that $\vecE(\vecr,t)\rightarrow \vecE(t)$ and
$\vecA(\vecr,t)\rightarrow \vecA(t)$, respectively,  the
electron-field interaction operator has the following forms in
length gauge  and velocity gauge:
\begin{equation}
H_{\mathrm{I}x}(t)=\left \{ \begin{array}{l}-e\vecr \cdot\vecE(t),\
(x=\mathrm{L})\\-\frac{e}{m} \hat\vecp \cdot \vecA(t) +\frac{e^2}{2m}
\vecA^2(t).\ (x=\mathrm{V}) \end{array} \right.\label{Hint}
\end{equation}
A free electron (no binding potential) in the presence of the laser
field is governed by the Hamiltonian \bq \Hf(t)
=\frac{\hat\vecp^2}{2m} +H_{\mathrm{I}x}(t). \eq The
time-evolution operator of the total Hamiltonian (\ref{totalH}) 
satisfies the Dyson
equation \bq U_x(t,t') = U_0(t,t')-i\int\limits^{t}_{t'}d\tau\,
U_x(t,\tau) H_{\mathrm{I}x}(\tau) U_0(\tau,t'),\label{dysonE} \eq
where $U_0(t,t')$ denotes the time-evolution operator of the
Hamiltonian (\ref{atomH}).

The exact ionization amplitude from an initial bound state
$|\psi_0(t)\rangle =|0\rangle \exp(i I_\mathrm{p} t)$ with ionization potential
 $I_\mathrm{p}$ to a
final continuum state  $|\psi_\vecp(t)\rangle$, both defined by the
Hamiltonian $H_0$, is  \bq M_\vecp = \lim_{t\to
\infty, t'\to -\infty} \langle
\psi_\vecp(t)|U_x(t,t')|\psi_0(t')\rangle. \label{mp} \eq We assume
that the laser field be turned off in the limits of $t\rightarrow
\infty$ and $t' \rightarrow -\infty$ and that
$\vecA(\infty)=\vecA(-\infty)=\mathbf{0}$. Gauge invariance then implies that
$M_\vecp$ be gauge invariant, and indeed this can easily be verified
explicitly. The SFA is obtained if we insert the Dyson
equation (\ref{dysonE}) into the ionization amplitude (\ref{mp}). 
The first term, which comes from
$U_0(t,t')$,  cancels since the initial and the final state are
orthogonal, and we are left with \cite{advances} \bq M_\vecp = -i \lim_{t \to
\infty} \int^t_{-\infty} d\tau \langle\psi_\vecp(t)|U_x(t,\tau)
H_{\mathrm{I}x}(\tau)|\psi_0(\tau)\rangle,\label{KFR0} \eq which is
still exact.

In the argument that follows we restrict ourselves for  the sake of
transparence and simplicity to ``direct'' electrons, i.e., those that
after the initial ionization never again feel the binding potential.
In order to obtain the transition amplitude for the direct electrons,
we replace in Eq.~(\ref{KFR0}) the exact state at time $\tau$, which
is $\langle \psi_\vecp(t)|U_x(t,\tau)$,  by the Volkov state
$\langle\VS_{\vecp x}(\tau)|$ [given below in Eq.~(\ref{VS})] where the
interaction with the binding potential is neglected. This yields the
well-known SFA amplitude \cite{KFR}
\begin{equation}
M_\vecp = -i  \int^\infty_{-\infty} d\tau \langle\VS_{\vecp x}(\tau)|
H_{\mathrm{I}x}(\tau)|\psi_0(\tau)\rangle. \label{KFR1}
\end{equation}
Here, for times $t < \tau$ the state of the electron is governed by
the Hamiltonian $H_0$, while for $t > \tau$ its time evolution
follows the Hamiltonian $\Hf$.

The matrix element (\ref{KFR1}) conveys the following physical
picture: for times $t< \tau$ the electron is sufficiently deeply
bound that to a good approximation its interaction with the laser
field can be ignored. At time $\tau$, it is ionized, and the laser
intensity is high enough to move the electron so rapidly out of the
range of the binding potential that now  the latter can be
neglected. 

However, this physical picture is in agreement with the formal
description only within L gauge. In L gauge, the interaction with
the laser field is accomplished by the scalar potential
$e\Phi(t)=H_{\mathrm{IL}}(t)$. There is no vector potential, so that
the operator of the velocity is $\hat\vecv=\vecp/m$. Hence,
$\hat\vecp^2/(2m)$ is the operator of the kinetic energy, and $H_0$
describes an atom that does not interact with the field, even if a
field is present. In V gauge, the operator of the velocity is
$\hat\vecv = [\hat\vecp -e\vecA(t)]/m$ where $\hat\vecp$ is the
operator of the canonical momentum. This is a conserved quantity
under the dipole approximation, but not a physical quantity
\cite{ct}, since $\hat\vecp = m\hat\vecv+e\vecA(t)$, and $\hat\vecv$
is a physical quantity while $\vecA(t)$ is not. In consequence, in
the presence of a laser field, the operator $H_0$ is not the
field-free Hamiltonian, and its eigenstate $|\psi_0(t_0)\rangle$
does incorporate some interaction with the field \cite{footnote1}.
Hence, in V gauge, the physical picture formulated above is not
supported by the matrix element (\ref{KFR1}).

It is instructive to evaluate the matrix element (\ref{KFR1}) by
the method of steepest descent, which is known to work very well for
sufficiently high intensity. We first recall the explicit form of
the Volkov wave function
\begin{equation}
\langle \vecr |\VS_{\vecp x}(t)\rangle =
\frac{e^{-iS_\vecp(t)}}{(2\pi)^\frac{3}{2}}
\left\{\begin{array}{l}e^{i\vecp\cdot\vecr},\ \ (x=
\mathrm{V})\\ e^{i(\vecp -e\vecA(t))\cdot\vecr},\
(x=\mathrm{L}) \end{array}\right. \label{VS}
\end{equation}
with the action
\begin{equation}
S_\vecp(t)= \frac{1}{2m}\int^t d\tau (\vecp-e\vecA(\tau))^2. \label{action}
\end{equation}
which has the same form in either
gauge \cite{footnote}.

Via an integration by parts, the transition amplitude (\ref{KFR1})
can be recast in the form \cite{advances}
\begin{equation}
M_\vecp = -i  \int^\infty_{-\infty} d\tau
\langle\VS_{\vecp x}(\tau)| V(\vecr)|\psi_0(\tau)\rangle,
\label{KFR2}
\end{equation}
which depends on the gauge only via the Volkov wave function
(\ref{VS}). Collecting the exponential time dependence of the
integrand in Eq.~(\ref{KFR2}) we find that the stationary points
with respect to $\tau$ are determined as the solutions of  the
saddle-point equation
\begin{equation}
[\vecp -e\vecA(\tau)]^2 = -2mI_\mathrm{p}. \label{speq}
\end{equation}
The transition amplitude then can be represented as the
superposition of the contributions of all those solutions $t_s$ of
Eq.~(\ref{speq}) for which  $\mathrm{Im} t_s >0$, with the result
\begin{equation}
M_\vecp = \sum_s V_{\vecp x s} \sqrt{\frac{2\pi i}{\vecE(t_s)\cdot[
\vecp-e\vecA(t_s)]}} e^{i[S_\vecp(t_s)+I_\mathrm{p} t_s]}.\label{mps}
\end{equation}
Only the form factor
\begin{equation}
V_{\vecp x s} =\left\{\begin{array}{l}\langle \vecp
|V(\vecr)|0\rangle,\ \  (x= \mathrm{V}),\\
\langle \vecp-e\vecA(t_s)|V(\vecr)|0\rangle,\ \ 
(x=\mathrm{L}) \end{array} \right.\label{ff}
\end{equation}
depends on the gauge. In V gauge, it is evaluated at the momentum
$\vecp$ at the detector, which is the same for all saddle-point
solutions. In L gauge, it is evaluated at the instantaneous velocity
at the ionization time $t_s$, whose component parallel to the laser
field according to Eq.~(\ref{speq}) is purely imaginary and can have
either sign. For a monochromatic linearly polarized laser field,
there are two solutions $t_s$ per cycle of the saddle-point equation
(\ref{speq}) with $\mathrm{Im}t_s>0$, one on either side of the
pertinent extremum of the vector potential.


To find out the signs of $\vecp-e\vecA(t_s)$ that correspond to the
solutions with $\mathrm{Im}t_s > 0$, let us consider the vector
potential $\vecA(t)=\vece A \cos \omega t$. We let $t=t_\mathrm{R}
+it_\mathrm{I}$, where $t_\mathrm{R}$ and $t_\mathrm{I}$ denote the
real and the imaginary part of $t$, respectively. The real and
imaginary parts of the saddle-point equation~(\ref{speq}) are
\begin{subequations}\label{realimag}
\begin{eqnarray}
p_\parallel-eA\cos \omega t_\mathrm{R} \cosh \omega t_\mathrm{I} = 0, \label{real}\\
eA \sin\omega t_\mathrm{R} \sinh \omega t_\mathrm{I} = 
\sqrt{2mI_\mathrm{p} +\vecp_\perp^2},\label{imag}
\end{eqnarray}
\end{subequations}
where $p_\parallel$ and $\vecp_\perp$ are the components of $\vecp$
parallel and perpendicular to the laser field and the square root may
have either sign. From
Eq.~(\ref{real}), the two solutions per cycle are such that $\cos
\omega t_\mathrm{R}$ has the same sign. Then, from Eq.~(\ref{imag})
and the fact that $t_\mathrm{I}>0$ for the physical saddle-point
solutions, we have that $\sin \omega t_\mathrm{R}$ has the opposite
sign for the two solutions. Hence, the two instantaneous velocities
that enter the L-gauge form factor (\ref{ff}) are
$\vecp-e\vecA(t_s)=(\pm i\sqrt{2mI_\mathrm{p}+\vecp_\perp^2},
\vecp_\perp)$. For $\vecp_\perp = \mathbf{0}$, they are purely
imaginary and have \textit{opposite} sign. This reflects the fact
that the electric field $\vecE(t_\mathrm{R})$ points in opposite direction
for the two solutions.

Now, for an even-parity ground state $|0\rangle$, $\langle
-\veca|V|0\rangle =\langle \veca|V|0\rangle$,  while for an
odd-parity state $\langle -\veca|V|0\rangle =- \langle
\veca|V|0\rangle$. Hence, for an odd-parity state, when in Eq.~(\ref{mps}) the
contributions of the two saddle points add in V gauge, they substract
in L gauge, and vice versa. 
Consequently, for an odd-parity initial state, constructive
interference in L gauge implies destructive interference in V gauge,
and vice versa. In contrast, for an even-parity ground state, both gauges
predict interference maxima and minima at the same positions. As soon as $\vecp_\perp \neq 0$, there is no
complete destructive or constructive interference anymore.

\begin{figure}
\includegraphics[width=0.46\textwidth]{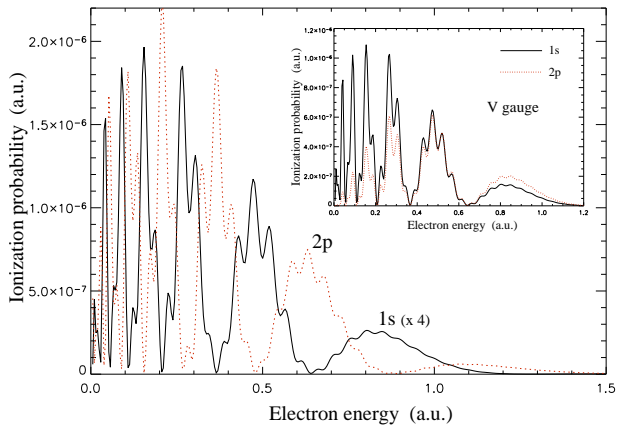}
\caption{(color online) SFA electron energy spectrum for emission in the (positive)
direction of the laser field (four-cycle $\sin^2$-pulse with $\phi=0$, $\omega=0.056$\, a.u., $E_0=0.0834$\,a.u.)  in L gauge, starting from an initial 1s (solid) or 2p (dashed) state.  The corresponding V gauge result is shown in the inset.}
\label{KFRnum}
\end{figure}

\begin{figure}
\includegraphics[width=0.46\textwidth,clip=true]{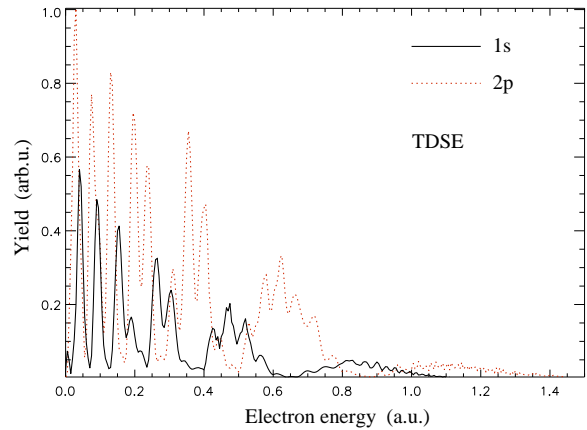}
\caption{(color online) Same as Fig.~\ref{KFRnum}, but computed from
the numerical solution of the TDSE.}
\label{TDSElow}
\end{figure}

\begin{figure}
\includegraphics[width=0.48\textwidth,clip=true]{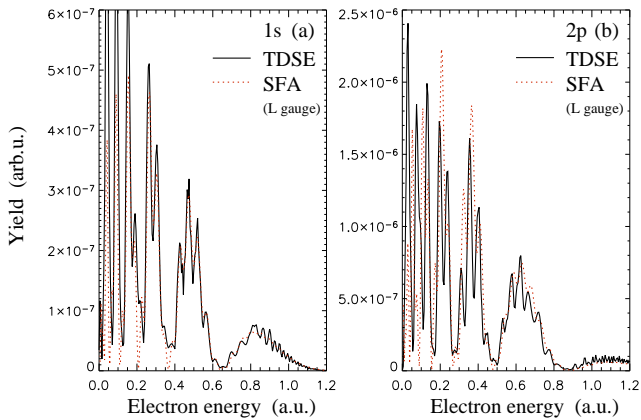}
\caption{(color online) Direct comparison of TDSE and SFA (L gauge) photoelectron spectra. The initial states are (a) $1s$ and (b) $2p$, the  other parameters as in Fig.\,\ref{KFRnum}. For convenience, the TDSE spectra are rescaled (but unshifted in energy). }
\label{COMP}
\end{figure}

In Figs.~\ref{KFRnum}, \ref{TDSElow}, and  \ref{COMP} we compare the results of
the SFA in L gauge and V gauge with a numerical solution of the
TDSE. All calculations have been carried out for a 4-cycle linearly
polarized laser pulse having the intensity $2.4\times
10^{14}\;\mathrm{W/cm}^2$ (field strength $E_0=$ 0.0834~a.u.), and
wavelength 800~nm (photon energy $\omega=$ 0.056~a.u.). The electric-field
vector is $E(t) \cos(\omega t+\phi) \hat{\mathbf{e}}$, with the
sine-square envelope $E(t)= E_0 \sin^2\frac{\omega t}{2n_\mathrm{p}}$
for $0\le t \le T_\mathrm{p}=n_\mathrm{p} T$, $T=2\pi/\omega$, and
$E(t)=0$ outside this interval. The carrier-envelope phase is $\phi=0$.
Figure~\ref{KFRnum} exhibits the results
of a numerical computation (not using the saddle-point
approximation) of the SFA amplitude (\ref{KFR1}) in L gauge and V
gauge, respectively, taking for $|\psi_0(t)\rangle$ the bound state of a zero-range potential \cite{GK,MGB03,GMB04,footnote2}. They illustrate the above statements. In other
words, in L gauge, everything else being equal, the positions of the
interference dips for a $p$ ground state coincide with those of the
interference humps for an $s$ ground state. In contrast, in V gauge
dips and humps occur at the same positions regardless of the parity
of the ground state. Figure~\ref{TDSElow} presents the corresponding
TDSE spectrum calculated by methods introduced elsewhere
\cite{DB05}. In order to mimic a short-range potential in the TDSE calculations, the Coulomb potential $-\Zeff/r$ has been cut at $r_\mathrm{c}=2$\,a.u. The nuclear charge $\Zeff$ was adjusted in such a way as to keep the ionization potential $I_\mathrm{p}$ at $0.5$\, a.u. for both the $1s$ and the $2p$ state. It has been shown recently \cite{DBJMO} that the agreement between SFA and TDSE low-energy electron spectra improves with decreasing potential range $r_\mathrm{c}$. A direct comparison of the TDSE and SFA (L gauge) results is presented in Fig.~\ref{COMP}. 
The agreement with respect to the energetic positions of the various peaks is  excellent. Residual discrepancies are observed in the shape of the spectrum  for low energies, especially for the $p$ ground state, and are likely due to the different large-distance behavior of the wave functions (zero range for the SFA vs. cut Coulomb for the TDSE).


The exact solution for ionization of negatively charged ions that is
available in the context of effective-range theory exhibits the interference
dips in complementary positions for  $s$ and $p$ ground states
\cite{manakov}, in agreement with the L-gauge SFA and the
exact TDSE solution.
The L-gauge SFA also appears to be supported by the experimental data:
the above-threshold-detachment energy spectrum for the negative
F$^-$ ion \cite{KH03}, which has a $p$ ground state, displays a
pronounced change of its slope at the energy where the L-gauge SFA
predicts an interference dip \cite{MGB03,GMB04}.

For elliptical polarization,  for ellipticities higher
than a  certain critical value the saddle-point equation
(\ref{speq}) only has one solution per cycle rather than two, so
that the interference ceases to exist \cite{PZWLBK}. This is so, in particular,
for circular polarization. Recently, the latter case was considered in
detail \cite{kiyan}. Even in the absence of interference, the form
factor (\ref{ff}) is still different in L gauge and in V gauge. For
an $s$ ground state $|0\rangle$, the form factor $\langle
\vecp|V|0\rangle$ has a maximum for $\vecp=\mathbf{0}$ and decreases with 
increasing $|\vecp|$, while for a
$p$ state, it has a zero at $\vecp=\mathbf{0}$ and extrema away from
$\vecp=\mathbf{0}$. In Ref.~\cite{kiyan}, for ionization of F$^-$ by
a circularly polarized laser field, the energy spectrum was
calculated in either gauge. The V-gauge spectrum peaks at a higher
energy than the L-gauge spectrum, which conforms with the
considerations given above. Moreover, Wigner's threshold law is only
reproduced in L gauge \cite{kiyan}.

Before concluding, we recall that in a numerical solution of the
TDSE the choice of gauge is ``merely'' a question of convenience. 
Generally, convergence is faster in V gauge where fewer angular momenta
contribute, much faster indeed for high intensity and low frequency \cite{CL}.
In contrast, in approximations such as the SFA, the choice of gauge is a 
contributing factor for the quality of the approximation. In fact, making
\textit{formally} the same approximation in two gauges may correspond
to different approximations \textit{physically}. A general argument in 
favor of the 
L gauge for use in the SFA has been put forward in Ref.~\cite{GK}: the L-gauge interaction Hamiltonian (\ref{Hint}) puts the emphasis on large distances from the atom, where the Volkov wave function is a good approximation to the final state. In a similar vein, we add that it appears to make more sense to evaluate
the form factor (\ref{ff}) at the instantaneous velocity at the time of ionization (as in L gauge) rather than at the drift velocity (as in V gauge), which for low frequencies the electron does not assume before it is far away from the ion. 

On the basis of a comparison with the solution of the time-dependent
Schr\"odinger equation, we conclude that the strong-field
approximation applied to above-threshold detachment of negative ions
affords a  better description in length gauge than in velocity
gauge. In view of the fundamental significance of the SFA for
strong-field physics, it is of great importance to find out which
gauge is better suited for above-threshold ionization of atoms and
molecules as well as nonsequential double ionization. In all of
these cases, the two gauges are known to yield different answers as
well.

We enjoyed discussions with M.\ Yu.\ Ivanov, H.\ G.\ Muller, and H.\ R.\ Reiss. This work was supported by VolkswagenStiftung, Deutsche
Forschungsgemeinschaft, and the NSERC Special Opportunity Program of Canada.


\begin{thebibliography}{99}
\bibitem{footnote1} This can be seen by calculating the
expectation value of the kinetic energy of the state
$|\psi_0(t)\rangle$. In V gauge, it is $\langle \psi_0(t)|(m/2)
\hat\vecv^2|\psi_0(t)\rangle = \langle \psi_0(t)|(m/2)[ \hat\vecp
-e\vecA(t)]^2|\psi_0(t)\rangle = \langle
\psi_0|\hat\vecp^2/(2m)|\psi_0\rangle + e^2\vecA^2(t)/(2m)$. We
assumed an eigenstate of parity so that the cross term vanishes.
Clearly, the kinetic energy depends on the field, so that
$|\psi_0(t)\rangle$ cannot be a field-free state. In order to obtain
a truly field-free state in  V gauge, we have to transform the
field-free state in L gauge to V gauge. This yields the state
$\exp[ie\vecA(t)\cdot\vecr]|\psi_0(t)\rangle$, which has been
called the ``noninteracting state in V-gauge'' \cite{SBBS}. If this
state were substituted in the transition amplitude (\ref{KFR2}) in
place of $|\psi_0(t_0)\rangle$, we would retrieve the same amplitude
(\ref{KFR2}) in L gauge. However, this is not what the derivation
presented above between Eqs.~(\ref{mp}) and (\ref{KFR1}) yields.
\bibitem{footnote} It may be irritating to see in L gauge, where
the vector potential is zero, the action  (\ref{action}) depend on
$\vecA(\tau)$. In this case, $\vecA(t)$ is used as short hand for
any function such that the electric field can be derived as
$\vecE(t)=-d\vecA(t)/dt$ while $\vecA(t)=\mathbf{0}$ outside the field.
\bibitem{footnote2} The dependence of the SFA results on the ground-state wave function is rather weak, in particular for an $s$ ground state and L gauge. This has been checked by redoing the calculations for hydrogenic wave functions.
\end{thebibliography}
\end{document}